\documentclass[mathlines]{aastex631}

\usepackage{amsmath}
\usepackage{subfigure}
\usepackage{textcomp}
\usepackage{xcolor}
\usepackage[misc]{ifsym} %% Letter symbol
\usepackage{hyperref}
\usepackage{threeparttable}
\usepackage{CJK}
\shortauthors{Tao et al.}
\graphicspath{{./}{figures/}}
\begin{document}
\begin{CJK*}{UTF8}{gbsn}
\title{A Deep SETI Search for Technosignatures in the TRAPPIST-1 System with FAST}

 \author[0009-0004-6992-1059]{Guang-Yuan Song}
 \altaffiliation{These authors contributed equally to this work.}
 \affiliation{College of Computer and Information Engineering, Dezhou University, Dezhou 253023, China}

 \author[0000-0002-4683-5500]{Zhen-Zhao Tao \href{mailto:taozhenzhao@dzu.edu.cn}{\textrm{\Letter}}}
 \altaffiliation{These authors contributed equally to this work.}
 \affiliation{College of Computer and Information Engineering, Dezhou University, Dezhou 253023, China;\url{taozhenzhao@dzu.edu.cn}}

 \author[0000-0002-8719-3137]{Bo-Lun Huang}
 \affiliation{Institute for Frontiers in Astronomy and Astrophysics, Beijing Normal University, Beijing 102206, China}

 %\author{Yan Cui \href{yancui@sdu.edu.cn}{\textrm{\Letter}}}
 %\affiliation{School of Law, Gansu University of Political Science and Law, Lanzhou 730070, China; \url{yancui@sdu.edu.cn}}

\author[0000-0002-9849-1762]{Bo Yu}
\affiliation{School of Mathematics and Big Data, Dezhou University, Dezhou 253023,  China}

\author[0000-0002-3363-9965]{Tong-Jie Zhang(张同杰) \href{mailto:tjzhang@bnu.edu.cn}{\textrm{\Letter}}}
\affiliation{Institute for Frontiers in Astronomy and Astrophysics, Beijing Normal University, Beijing 102206, China}
\affiliation{School of Physics and Astronomy, Beijing Normal University, Beijing 100875, China; \url{tjzhang@bnu.edu.cn}}
\affiliation{Institute for Astronomical Science, Dezhou University, Dezhou 253023, China}
\begin{abstract}

The Five-hundred-meter Aperture Spherical Telescope (FAST) is the world's largest single-dish radio telescope, and the search for extraterrestrial intelligence (SETI) is one of its five key science objectives. 
%Its exceptional sensitivity in the low-frequency L band makes it a powerful instrument for technosignature searches.
We conducted a targeted search for unresolved narrowband radio technosignatures toward the TRAPPIST-1 system using FAST. The observations consisted of five independent L-band pointings, each with a 20-minute integration, for a total on-source time of 1.67\,h. The frequency coverage spanned 1.05--1.45\,GHz with a spectral resolution of $\sim$7.5\,Hz. We searched Stokes I (total intensity) data for drifting signals with Doppler drift rates within $\pm 20\,\mathrm{Hz\,s^{-1}}$ and a signal-to-noise ratio threshold of $\mathrm{S/N}=10$.
For unresolved, minimally drifting narrowband signals, the nominal minimum detectable equivalent isotropic radiated power is $\mathrm{EIRP}_{\min}\approx1.44\times10^{10}\,\mathrm{W}$. We explicitly quantify how this limit degrades at high drift rates because of tree de-Doppler power smearing, reaching $\approx3.84\times10^{11}\,\mathrm{W}$ at $|\dot{\nu}|=20\,\mathrm{Hz\,s^{-1}}$ under a conservative peak-channel sensitivity correction.
No credible technosignature candidates attributable to the TRAPPIST-1 system were identified within the searched parameter space. Nevertheless,
%with its seven Earth-sized planets, three of which reside in the habitable zone, 
TRAPPIST-1 remains a compelling target for future SETI efforts. We plan to extend our search to other signal types, such as periodic, transient, broadband, and propagation-broadened transmitters, and to carry out broader surveys of nearby exoplanetary systems with FAST.

\end{abstract}

\keywords{\href{http://astrothesaurus.org/uat/74}{Astrobiology (74)}; \href{http://astrothesaurus.org/uat/2127}{Search for extraterrestrial intelligence (2127)}; \href{http://astrothesaurus.org/uat/2128}{Technosignatures (2128)}; \href{http://astrothesaurus.org/uat/498}{Exoplanets (498)}}

%% We recommend that authors also use the natbib  \citeppp
%% and  \citept commands to identify citations.  The citations are

\section{Introduction} \label{sec:intro}

The search for life beyond Earth—and particularly for intelligent life—represents one of humanity’s most profound scientific and philosophical pursuits. The Search for Extraterrestrial Intelligence (SETI) seeks to detect technosignatures: observational evidence of technology that cannot be explained by natural astrophysical processes \citep{2001ARA&A..39..511T}. Among the most compelling of these are narrowband radio signals (with bandwidths on the order of hertz). This narrowness is a key criterion in radio SETI: while the narrowest naturally occurring astrophysical emitters, such as masers, are limited by thermal and turbulent broadening to line widths of several hundred hertz, no known natural process produces signals only a few hertz wide \citep{1997ApJ...487..782C}. Such signals, being energetically efficient and precisely tuned, would strongly suggest an artificial origin \citep{1959Natur.184..844C}.

Over the past two decades, several large-scale SETI programs have advanced the search for narrowband radio technosignatures. SETI@home, launched in 1999, pioneered the use of distributed volunteer computing to analyze Arecibo telescope data for candidate signals \citep{2000ASPC.Cobb.sh,2025AJ..Korpela.sh,2001CSE..Korpela.sh,2025AJ..Anderson.sh}. More recently, the Breakthrough Listen initiative has conducted the most comprehensive and sensitive SETI survey to date, using facilities such as the Green Bank Telescope and the Parkes Murriyang radio telescope to scan millions of narrow frequency channels across thousands of nearby stars and galaxies \citep{2017ApJ...849..104E,2020AJ....159...86P,2021AJ....162...33G,2023NatAs.bl,  2025AJ..Pardo.bl}. %\citep{2013ApJ...767...94S,2017ApJ...849..104E,2019AJ....157..122P,2020AJ....159...86P,2020AJ....160...29S,2021AJ....161..286T,2021AJ....162...33G,2021NatAs.tmp..203S,2023NatAs.bl, 2023MNRAS.Uno.bl, 2025AJ..Pardo.bl,2023AJ..Sheikh.bl,2024AJ...Choza.bl,2024AJ..Brzycki.bl,2023AJ..Acharyya.bl,2023AJ.Zuckerman.bl,2025AJ...Painter.bl,2023MNRAS.Garrett.bl,2025PASA...Barrett.bl,2024yCat..Zuckerman.bl}. 
These efforts have refined search algorithms, established robust radio frequency interference (RFI) mitigation strategies, and set benchmarks for survey speed and sensitivity.

%Building upon these foundations, FAST’s extraordinary collecting area enables us to probe even fainter transmitters than previously possible, opening new parameter space for the detection of weak or transient technosignatures.
Detecting such faint signals over interstellar distances demands instruments of exceptional sensitivity. The Five-hundred-meter Aperture Spherical radio Telescope (FAST) is currently the world’s largest single-dish radio telescope, offering unmatched collecting area and sensitivity for single-dish observations. These capabilities make FAST a leading facility for advancing the frontiers of SETI research \citep{2011IJMPD..20..989N}. Since the start of science operations, FAST has executed several pioneering SETI programs, beginning with the first FAST SETI survey, which established a foundational data-processing pipeline and demonstrated a commensal observing strategy \citep{2020ApJ...891..174Z}. Subsequent campaigns have expanded to deep, targeted observations of nearby stars and known exoplanetary systems, delivering robust methodologies and setting new sensitivity benchmarks in the field \citep{2022AJ....164..160T,2023AJ....165..132L,2023AJ..tao,2023AJ..huang,2025AJ..luan}.

\begin{figure}[ht!]
\centering
\includegraphics[width=0.8\columnwidth]{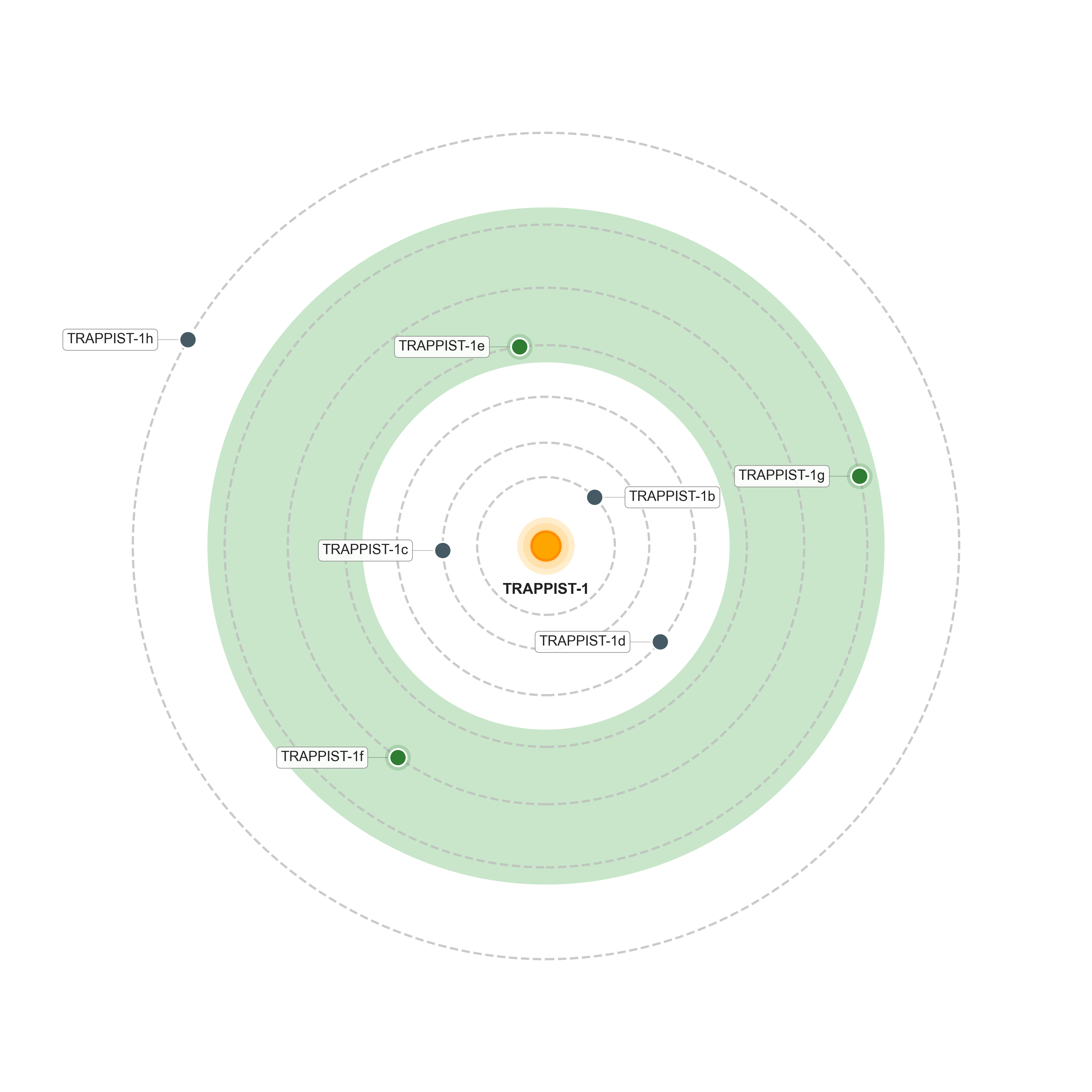} % <-- 确保文件名正确
\caption{Schematic of the TRAPPIST-1 planetary system. The seven known planets are shown orbiting the central star \citep{2017Natur.Gillon,2021PSJ..Agol}. The semi-transparent green annulus marks the optimistic habitable zone, which contains planets e, f, and g, which are the primary targets of this search \citep{2020SSRv.Turbet}.}
\label{fig:system}
\end{figure}

% In this work, we turn FAST’s capabilities toward one of the most intriguing planetary systems discovered to date: TRAPPIST-1. This nearby ultracool dwarf hosts seven transiting, Earth-sized planets, at least three of which—TRAPPIST-1e, f, and g—orbit in the star’s optimistic habitable zone, where liquid water could potentially exist on their surfaces \citep{2017Natur.Gillon}. The system’s compact orbital architecture is illustrated in Figure~\ref{fig:system}. Its combination of terrestrial-sized planets, proximity, and favorable conditions makes TRAPPIST-1 a keystone target for both astrobiology and technosignature searches.
TRAPPIST-1 is one of the most compelling targets for both astrobiology and technosignature searches. This nearby ultracool dwarf hosts seven known transiting, Earth-sized planets, at least three of which (TRAPPIST-1\,e, f, and g) reside within the star's optimistic habitable zone where liquid water could potentially exist on their surfaces \citep{2017Natur.Gillon,2020SSRv.Turbet}. The system's compact, coplanar architecture (Figure~\ref{fig:system}), proximity, and frequent transits make it an exceptional laboratory for studies of planetary environments and for searches targeting leaked emission or directed technosignatures. In particular, \citet{2024AJ.Tusay} carried out the longest single-target radio technosignature search of TRAPPIST-1 to date using the Allen Telescope Array, acquiring roughly 28\,h of beamformed data spanning $\sim$0.9--9.3\,GHz. They implemented the \texttt{NbeamAnalysis} filtering pipeline and targeted narrowband signals coincident with predicted planet-planet occultations (PPOs), reporting non-detections and EIRP upper limits for PPO events of order a few to a few tens of terawatts for minimally drifting signals \citep{2024AJ.Tusay}. While that wideband, long-dwell survey provides important constraints on high-power transmitters across a broad frequency range, the extreme collecting area of FAST enables complementary searches that probe much lower transmitted powers in the L band.

% The unusual compactness of TRAPPIST-1 motivates search strategies beyond the classic model of deliberate, omnidirectional beacons. Here, we explore the interplanetary communication hypothesis: the light-travel time between adjacent planets is only a few minutes, making it plausible that a technologically advanced civilization (or multiple civilizations) could maintain a local network for communication, navigation, or trade. Such interplanetary transmissions would require far less power than interstellar beacons and could represent a more common form of leakage technosignature\citep{2023ApJ..Ashtari,2024AJ.Tusay}.
% The system also provides a fortuitous geometric advantage. All seven planets are in near-perfect coplanar orbits \citep{2017NatAs.Luger}, viewed almost exactly edge-on from Earth—an alignment that enables all transits to be observed \citep{2017Natur.Gillon}. This orientation could act as a natural “funnel” for any radio traffic confined to the orbital plane, increasing the odds of intercepting signals not intended for us.

In this paper, we present the results of a deep FAST SETI search targeting TRAPPIST-1. In \S\ref{sec:obs}, we describe the observational setup and data analysis pipeline, highlighting the use of Multi-Beam Coincidence Matching (MBCM) for robust RFI rejection. In \S\ref{sec:results}, we present the search outcomes and quantify how transmitter-power limits depend on receiver sensitivity, drift rate, and apparent signal bandwidth. In \S\ref{sec:discussion}, we assess the completeness of our search for different classes of signals (e.g., low-duty-cycle, transient, broadband, propagation-broadened, frequency-agile, and highly directed emissions) and consider implications for future SETI strategies; finally, in \S\ref{sec: Conclusions} we summarize our findings and outline planned follow-up observations.
%we extend these constraints to the case of periodic or intermittent signals and discuss the implications for future, higher-cadence or flare-triggered SETI strategies, before concluding in \S\ref{sec: Conclusions}.

\section{Observations and Data Analysis}
\label{sec:obs}
The primary scientific goal of this work was to conduct a deep search for unresolved narrowband radio technosignatures from the TRAPPIST-1 system, motivated by its unique architecture featuring multiple habitable-zone planets. A key challenge in any radio SETI search is the mitigation of RFI. We employed the MBCM strategy introduced above, which performs simultaneous ON-target and OFF-target observations to effectively identify and reject RFI \citep{2022AJ....164..160T,2023AJ....165..132L,2023AJ..tao,2025AJ..luan}.

\begin{figure}[ht!]
\centering
\includegraphics[width=0.6\columnwidth]{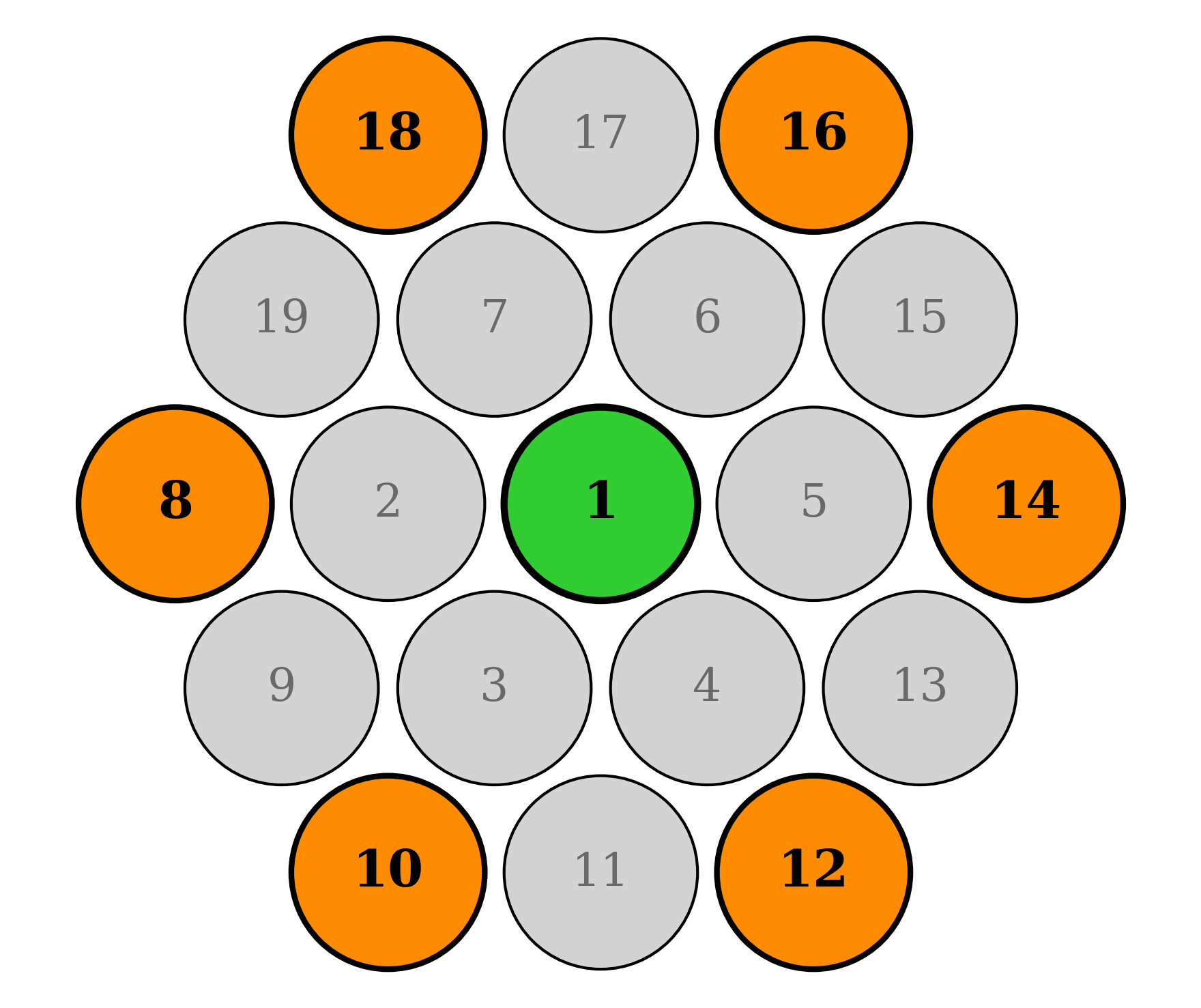} % <-- 替换为您的波束图文件名
\caption{Schematic diagram of the FAST 19-beam receiver layout on the sky, illustrating the physical configuration for the MBCM strategy. The central beam (1, green) serves as the ON-target beam, while the six outermost beams (orange) are used as simultaneous OFF-target references. The remaining beams are shown in gray. This physical configuration is adapted from \cite{2023AJ..tao}.}
\label{fig:beam_layout}
\end{figure}

This strategy utilizes the FAST L-band 19-beam receiver. As illustrated in Figure~\ref{fig:beam_layout}, the central beam (Beam~1) was configured to track TRAPPIST-1 continuously, serving as the ON-target observation. The six outermost beams (Beams 8, 10, 12, 14, 16, and 18) were simultaneously recorded as OFF-target references. Prior FAST MBCM work quantified that a celestial point source in the central beam is strongly attenuated in these widely separated reference beams, whereas local and near-Earth RFI can illuminate multiple beams or enter through sidelobes \citep{2022AJ....164..160T,2023AJ..tao,2025AJ..luan}. This single-dish multibeam response differs from the sparse-array beamformed case discussed by \citet{2024AJ.Tusay}, but both approaches use spatial localization to distinguish distant sky-localized emission from human-made interference.
% ===================================================================
% Figure 2: Beam Layout
% ===================================================================
% ===================================================================
Our observations were conducted in five sessions between 2024 October 3 and 2024 October 27, as detailed in Table~\ref{tab:obslog}. The original observing logs recorded local Beijing Time (UTC+8); Table~\ref{tab:obslog} reports the corresponding UTC start times and Modified Julian Dates. Each session consisted of a 20-minute (1200~s) tracking observation, resulting in a total on-source integration time of 100~minutes (1.67~hours). The intervals between observations varied non-uniformly from 5 to 7 days. This quasi-weekly cadence over a 24-day baseline was intentionally designed to provide phase coverage of the short orbital periods of the habitable-zone planets and to reduce aliasing with potential periodic human-made signals (further discussed in \S\ref{sec:discussion}).

Data were collected over a frequency range of 1.0--1.5~GHz using the FAST SETI backend.\footnote{FAST FRB/SETI backend documentation: \url{https://casper.berkeley.edu/wiki/FRB/SETI_backend_for_FAST_19-horn_receiver}.} The recorded data have 65,536k frequency channels, corresponding to a fine frequency resolution of $\sim$7.5~Hz. Each spectrum was integrated for 10~s. The raw data were stored in FITS format \citep{1981A&AS...44..363W}, containing four polarization products (XX, YY, XY, YX). As the top and bottom 50~MHz of the band are outside the receiver's optimal frequency range, our effective analyzed bandwidth is 400~MHz, from 1050~MHz to 1450~MHz.

% --- Table 1 (Observation Log) should be placed here ---
\begin{deluxetable}{ccccccc}[ht!]
\tablecaption{FAST Observation Log for TRAPPIST-1 \label{tab:obslog}}
\tablehead{
\colhead{Obs. ID} & \colhead{Date} & \colhead{Start Time} & \colhead{MJD\tablenotemark{a}} & \colhead{ZA\tablenotemark{b}} & \colhead{Center Frequency} & \colhead{Bandwidth} \\
\colhead{} & \colhead{(YYYY-MM-DD)} & \colhead{(UTC)} & \colhead{(day)} & \colhead{(deg)} & \colhead{(GHz)} & \colhead{(GHz)}
}
\startdata
1 & 2024-10-03 & 15:46 & 60586.657 & 32.62 & 1.25 & 0.4 \\
2 & 2024-10-09 & 15:13 & 60592.634 & 31.89 & 1.25 & 0.4 \\
3 & 2024-10-16 & 14:32 & 60599.606 & 31.11 & 1.25 & 0.4 \\
4 & 2024-10-21 & 13:46 & 60604.574 & 30.56 & 1.25 & 0.4 \\
5 & 2024-10-27 & 14:42 & 60610.613 & 35.88 & 1.25 & 0.4 \\
\enddata
\tablecomments{All observations were conducted using the L-band 19-beam receiver on FAST. Each observation had a duration of 1200~s (20~minutes). The UTC times shown here were converted from Beijing Time (UTC+8) recorded in the observing logs.}
\tablenotetext{a}{Modified Julian Date at the start of the observation, rounded to 0.001~d to match the minute-level timing precision.}
\tablenotetext{b}{Zenith angle at the midpoint of each 20-minute observation, calculated using the FAST site coordinates and the TRAPPIST-1 coordinates.}
\end{deluxetable}
% ----------------------------------------------------
The raw data from the central beam (Beam 1) and the six outermost reference beams (Beams 8, 10, 12, 14, 16, and 18) were processed to search for narrowband, drifting signals.
%The raw data from all 19 beams were processed to search for narrowband, drifting signals. 
The logical pipeline used to filter these signals and identify candidates is shown in Figure~\ref{fig:flowchart}. In our pipeline, we define two distinct classes of signals:

% ===================================================================
% Figure 3: Flowchart
% ===================================================================
\begin{figure}[ht!]
\centering
\includegraphics[width=0.8\columnwidth]{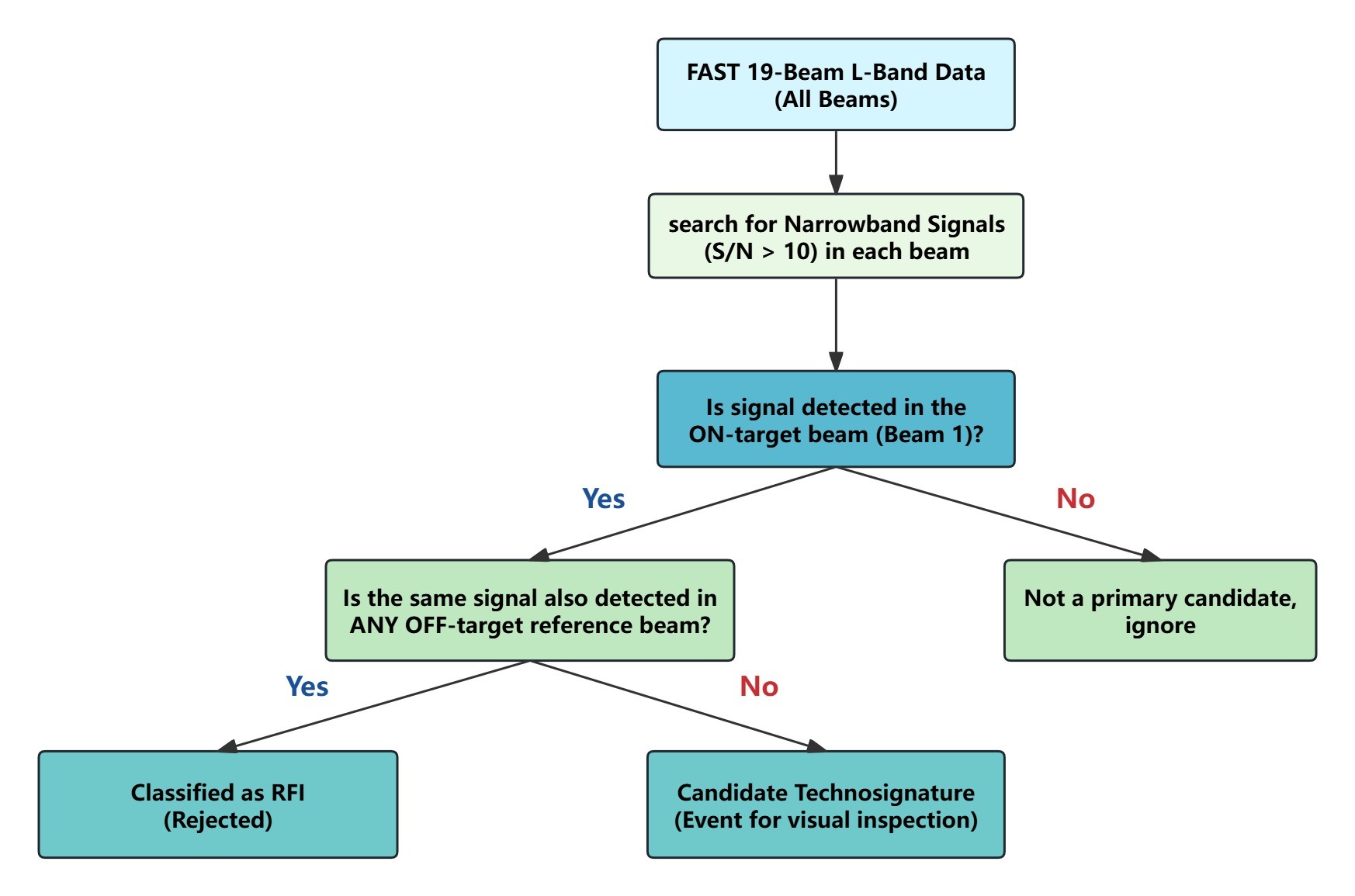} % <-- 替换为您的流程图文件名
\caption{A logical flowchart of the MBCM pipeline used for RFI mitigation. A signal is only classified as a candidate technosignature (event) if it is detected in the central ON-target beam and is simultaneously absent from all six OFF-target reference beams. In this search, \texttt{turboSETI} returned 35,398 hits in the seven analyzed beams; the MBCM filter reduced these to 94 events, all of which were rejected as human-made RFI after visual inspection.}
\label{fig:flowchart}
\end{figure}
% ===================================================================  
\begin{itemize}
    \item \textbf{Hit:} A narrowband signal detected in any of the 7 analyzed beams at the adopted signal-to-noise ratio threshold of S/N = 10.
    %A narrowband signal detected in \textit{any} of the 19 beams with a signal-to-noise ratio (S/N) exceeding 10.
    \item \textbf{Event:} A hit that passes the MBCM filter; i.e., it is detected in the ON-target beam (Beam~1) but is absent from all six OFF-target reference beams.
\end{itemize}

The core of our search pipeline is the \texttt{turboSETI} software package, which employs a tree de-Doppler algorithm to search for drifting signals \citep{2017ApJ...849..104E, 2019ascl.soft06006E}. The FITS files from each observation were first converted into filterbank format. We generated Stokes I spectra by summing the XX and YY polarization data, and then searched this total intensity data for candidate signals.
%We then searched the data from both XX and YY polarizations independently.

%A narrowband signal from a distant, accelerating source will exhibit a frequency drift over time ($\dot{\nu}$). This drift rate is a key search parameter. Given that the planets in the TRAPPIST-1 system are in tight orbits, we expect potentially significant accelerations. 
%While the innermost planets could theoretically produce very high drift rates, our search was primarily motivated by the system's three planets residing in the optimistic habitable zone (TRAPPIST-1e, f, and g). Due to their longer orbital periods, the maximum line-of-sight accelerations originating from these primary targets are considerably lower. 
%Therefore, we adopted a drift rate range of ±20 Hz s⁻¹. While this range is significantly wider than the accelerations expected from the habitable zone planets, it allows for the detection of potential signals with higher drift rates originating from more extreme orbital configurations or other artificial sources.
%Therefore, we adopted a drift rate range of $\pm$4~Hz~s$^{-1}$, a range that fully encompasses the theoretically predicted drift rates for these principal habitable zone planets while maintaining computational efficiency.
%Consistent with the capabilities of the FAST L-band backend and previous searches,
%we set the maximum drift rate (MDR) to $\pm$20~Hz\,s$^{-1}$ and the S/N threshold to 10. \texttt{turboSETI} outputs a DAT file for each beam containing a list of all detected hits.
A narrowband signal emitted by a stable transmitter will exhibit a frequency drift over time ($\dot{\nu}$) when there is a relative line-of-sight acceleration between the transmitter and the receiver, including contributions from both the source system and the motion of Earth \citep{2013ApJ...767...94S,2023AJ....166..182L}. The compact architecture of TRAPPIST-1 makes drift-rate coverage important, but the relevant acceleration depends on the assumed transmitter location and motion rather than on compactness alone. For an order-of-magnitude check, the maximum orbital acceleration scale for a surface transmitter corotating with a planet is $a_{\rm orb}=4\pi^{2}a/P_{\rm orb}^{2}$, giving an approximate drift-rate scale $\dot{\nu}\simeq\nu a_{\rm orb}/c$. At the top of our analyzed band ($\nu=1.5$~GHz), this gives $\sim$3.1, 1.8, and 1.2~Hz\,s$^{-1}$ for the habitable-zone planets e, f, and g, respectively, using the orbital elements from \citet{2021PSJ..Agol}. The innermost planet b gives a scale of $\sim$20~Hz\,s$^{-1}$. We therefore adopted a broad drift-rate range of $\pm 20$~Hz\,s$^{-1}$, which comfortably covers the expected surface-transmitter drift rates for the principal habitable-zone targets and reaches the approximate inner-planet orbital-acceleration scale. This range also covers some orbiting transmitters, but it does not encompass all possible extreme satellite or spacecraft configurations. Non-synchronous or chaotic spin states would expand the possible drift-rate distribution. The S/N threshold was set to 10, a standard value in many narrowband SETI searches that balances candidate recovery against the large number of time-frequency trials \citep{2017ApJ...849..104E,2020AJ....159...86P,2024AJ.Tusay}. \texttt{turboSETI} generates a DAT file for each beam containing a list of all detected hits.

Finally, a modified version of the \texttt{find\_event\_pipeline} from \texttt{turboSETI} was used to perform the coincidence matching. This script cross-references the hits from the ON-target beam with those from the six OFF-target beams and outputs a CSV file containing only the events that are unique to the TRAPPIST-1 pointing. The modifications are limited to matching the FAST seven-beam MBCM configuration and the adopted event-output format. These events are the final candidates for visual inspection. The modified code is not yet maintained as a public software release; it will be made available upon reasonable request, and the logic is documented here to make the event definition reproducible. We validated the matching step by confirming that synthetic hit lists with known ON-only and multi-beam RFI-like entries were classified as expected before applying the code to the TRAPPIST-1 data. This validation tests the coincidence-matching logic rather than the end-to-end detectability of injected signals in raw data.

\section{Results}
\label{sec:results}

Our data processing pipeline was applied to the Stokes I (total intensity) data. The initial search with \texttt{turboSETI} yielded a total of 35,398 hits. After applying the MBCM filter, these numbers were drastically reduced, leaving only 94 events for further vetting.
%Our data processing pipeline was applied independently to the two orthogonal linear polarizations (XX and YY). The initial search with \texttt{turboSETI} yielded a total of 238,672 hits for the XX polarization and 236,590 for the YY polarization. After applying the Multi-Beam Coincidence Matching (MBCM) filter, these numbers were drastically reduced, leaving only 707 events for the XX polarization and 844 for the YY polarization for further vetting.

%The statistical distributions of these signals for both polarizations are presented in Figure~\ref{fig:stats_xx} and \ref{fig:stats_yy}. A key finding is the high degree of similarity between the XX and YY distributions. The general trends for frequency, drift rate, and S/N do not vary significantly between the two polarizations, which strongly indicates that the detected signals, overwhelmingly dominated by RFI, are not significantly polarized. 

\begin{figure*}[ht!] 
\centering
\includegraphics[width=\textwidth]{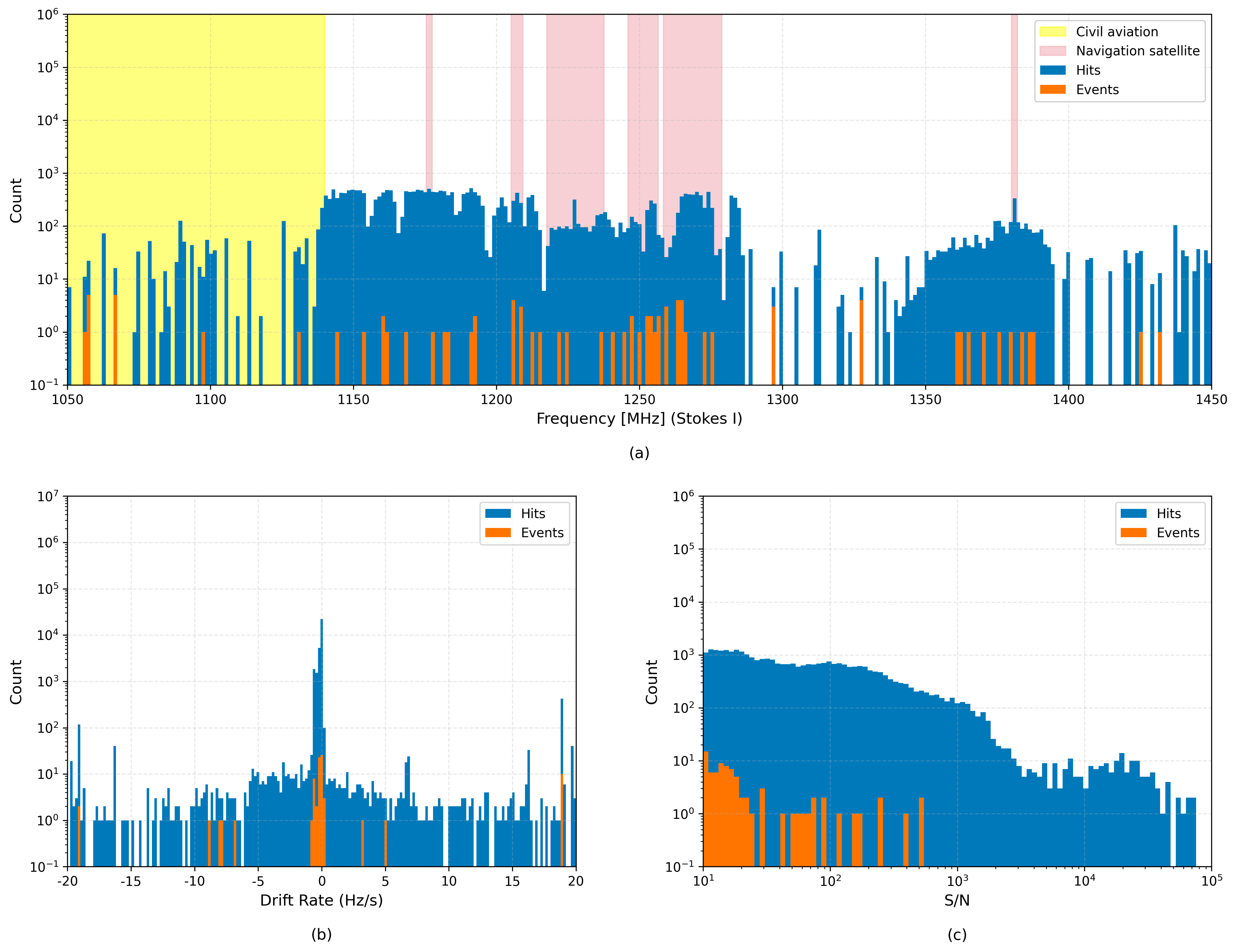} % <-- 替换为XX图的文件名
\caption{The blue histograms represent the distribution of all hits found in the 7 analyzed beams (the central beam and six reference beams). The orange histograms represent the events that passed the MBCM filter. Panels show: (a) Signal distribution across frequency for Stokes I, with known RFI sources shaded. (b) Signal distribution by Doppler drift rate. (c) Signal distribution by signal-to-noise ratio (S/N).}
\label{fig:stats_i}
\end{figure*}

As seen in Figure~\ref{fig:stats_i}(a), some known RFI bands contain significant concentrations of hits, although many hits also occur outside these shaded bands. According to RFI environment tests at the FAST site, the shaded bands are primarily associated with civil aviation and navigation satellites \citep{2021RAA....21...18W}. While only a small fraction of hits fall within the civil aviation band (1030--1140~MHz), a considerable proportion are found within the various navigation satellite bands (e.g., 1176.45$\pm$1.023~MHz, 1227.6$\pm$10~MHz, etc.).
The drift rate distributions (Figure~\ref{fig:stats_i}(b)) show a strong peak at 0~Hz\,s$^{-1}$, as expected from stationary, ground-based RFI sources. The slight bias towards negative drift rates is consistent with signals from non-geosynchronous satellites and other moving human-made sources. Furthermore, the majority of events have low S/N ratios (Figure~\ref{fig:stats_i}(c)). This is a common outcome of the MBCM method, where weak RFI signals may accidentally fall below the detection threshold in the reference beams while remaining above it in the on-target beam, thus passing the filter.

\begin{figure*}[ht!] 
\centering
\includegraphics[width=0.6\columnwidth]{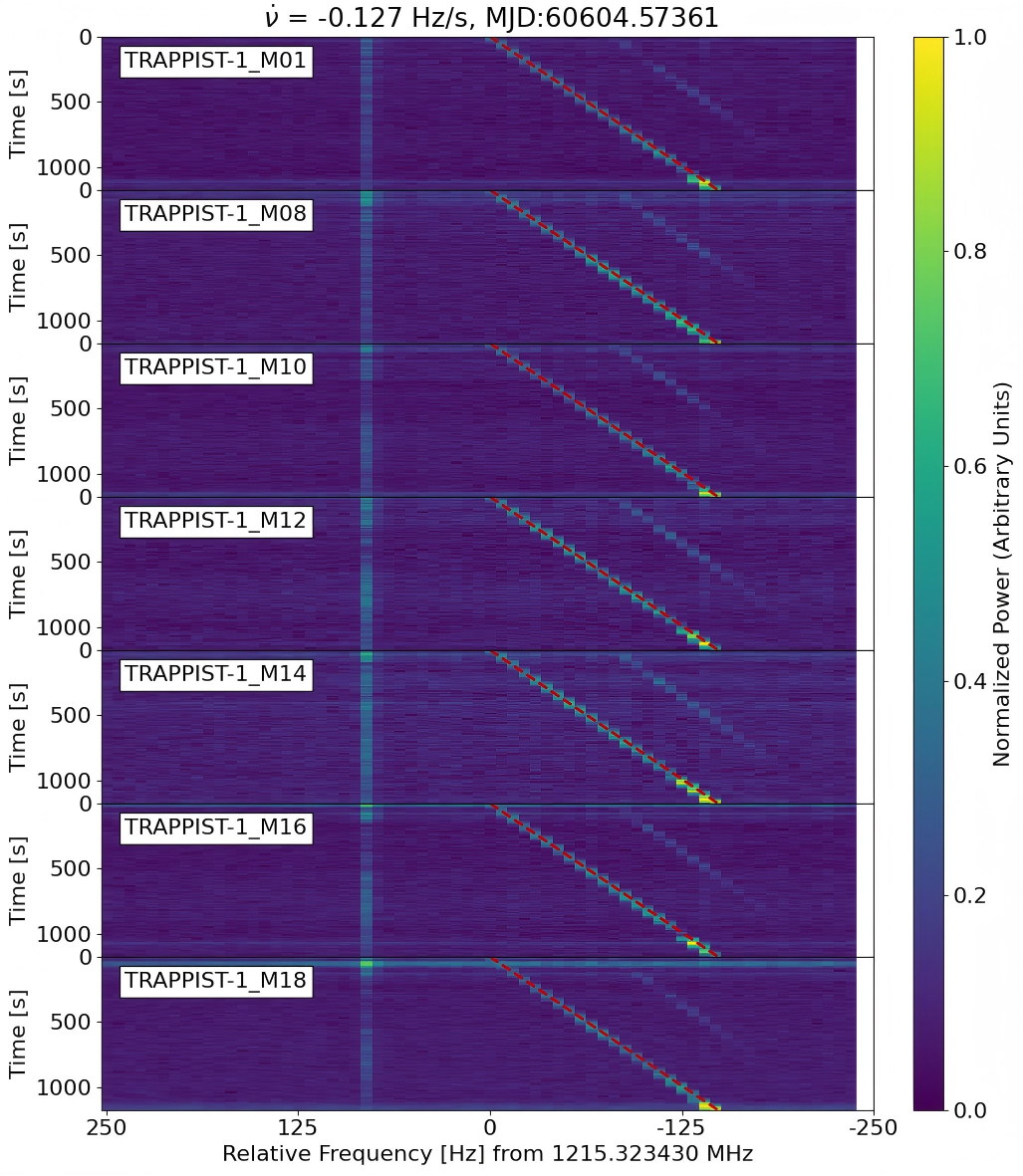} % <-- 替换为YY图的文件名
\caption{Example of a false-positive candidate signal identified as RFI through visual inspection of dynamic spectra (waterfall plots). The top panel displays the spectrum from the ON-target beam (Beam 1) pointing toward TRAPPIST-1, while the lower panels show the simultaneous data from the six OFF-target reference beams (Beams 8, 10, 12, 14, 16, and 18). The signal, detected at $\sim$1215.32 MHz, exhibits a \texttt{turboSETI}-reported drift rate of $-0.127$ Hz s$^{-1}$, indicated by the red dashed guide line. Each panel is normalized independently for visual inspection; therefore the color scale should be interpreted as morphology within a panel rather than as a calibrated beam-to-beam power comparison. The concurrent appearance of the same drifting morphology in multiple reference beams, even where it did not pass the automated S/N threshold, identifies the event as human-made RFI rather than a localized TRAPPIST-1 signal.}
\label{fig:ex1}
\end{figure*}

All 94 events were subjected to rigorous visual inspection of their dynamic spectra (waterfall plots) using the \texttt{BLIMPY} package \citep{2019JOSS....4.1554P}. We rejected events when it was clear by eye that a signal was present in the reference beams, even if it was below the nominal S/N threshold. This final vetting step revealed that all events could be attributed to RFI.
Figure~\ref{fig:ex1} illustrates a representative example of a candidate signal identified during the initial search but subsequently rejected by our filtering pipeline. In the top panel (ON-target Beam 1), the signal appears as a distinct, narrowband track with a non-zero drift rate, mimicking the characteristics of a potential technosignature. However, the simultaneous presence of similar signal structures in the OFF-target reference beams (lower panels) confirms a human-made local or near-Earth origin. This demonstrates the critical role of the MBCM strategy in distinguishing localized celestial sources from widespread RFI.
Consequently, after a comprehensive search and multi-stage vetting process, no credible technosignature candidates attributable to the TRAPPIST-1 system survived the full vetting process in the total intensity data within the sensitivity limits and parameter space of our survey.
%we report a null detection of persistent or periodic radio technosignatures in either polarization within the sensitivity limits and parameter space of our survey.

% --- Figure 4: XX Polarization Stats ---
% -----------------------------------------------------------------------------------

% --- Figure 5: YY Polarization Stats ---
% -----------------------------------------------------------------------------------
% -----------------------------------------------------------------------------------

From our null result, we can place sensitivity limits on the minimum detectable equivalent isotropic radiated power (EIRP$_{\rm min}$) of potential transmitters within the TRAPPIST-1 system. These limits apply directly to unresolved narrowband signals whose apparent bandwidth is no larger than one analyzed frequency channel and whose emission persists over a 1200~s pointing. The EIRP$_{\rm min}$ is defined as:
\begin{equation}
    \text{EIRP}_{\text{min}}=4\pi d^{2} S_{\text{min}},
\label{eq:eirp}
\end{equation}

where $d$ is the distance to TRAPPIST-1, which we take to be $12.5\ \mathrm{pc}$ based on Gaia astrometry \citep{2021A&A...649A...1G}; $S_{\text{min}}$ is the minimum detectable flux. For an unresolved narrowband signal, $S_{\text{min}}$ is given by \citet{2017ApJ...849..104E}:
\begin{equation}
    S_{\text{min}} = \sigma_{\text{min}}~ \frac{2k_{\text{B}}T_{\text{sys}}}{A_{\text{eff}}} \sqrt{\frac{\delta\nu}{n_{\text{pol}} ~ t_{\text{obs}}}},
\label{eq:smin}    
\end{equation}
where $\sigma_{\text{min}}$ is the S/N threshold, $k_{\text{B}}$ is the Boltzmann constant, and the ratio $A_{\text{eff}}/T_{\text{sys}}$ represents the sensitivity of the telescope.

The FAST L-band 19-beam receiver has a reported central-beam sensitivity of order $A_{\text{eff}}/T_{\text{sys}}\sim2000$~m$^{2}$\,K$^{-1}$, with frequency and zenith-angle dependence across the L band \citep{2011IJMPD..20..989N,2016RaSc...51.1060L,2020RAA....20...64J}. Our observations occurred at midpoint zenith angles of 30.56--35.88 deg (Table~\ref{tab:obslog}), outside the highest-sensitivity low-ZA regime but within the nominal FAST observing range. Because contemporaneous calibrator-derived values of $A_{\text{eff}}/T_{\text{sys}}$ are not available for these exact pointings, we adopt $2000$~m$^{2}$\,K$^{-1}$ as the nominal value and bracket the calculation with a conservative 1800--2200~m$^{2}$\,K$^{-1}$ range to account for the frequency- and ZA-dependent fluctuations reported by \citet{2020RAA....20...64J}.
We use $\sigma_{\text{min}}=10$, $\delta\nu \approx 7.5$~Hz, $n_{\text{pol}}=2$ for Stokes I, and $t_{\text{obs}}=1200$~s. Plugging these nominal values into Equation~\ref{eq:smin}, we calculate a minimum detectable flux of:
\begin{equation}
    S_{\text{min}} = 7.7 \times 10^{-27}~\text{W\,m}^{-2}
\end{equation}

Using this value for $S_{\text{min}}$ in Equation~\ref{eq:eirp}, we then calculate the nominal minimum detectable EIRP limit for unresolved, minimally drifting narrowband signals:
\begin{equation}
    \text{EIRP}_{\text{min}} = 1.44 \times 10^{10}~\text{W}
\end{equation}

The adopted sensitivity bracket changes this nominal value to the range shown in Table~\ref{tab:sensitivity_range}.

\begin{deluxetable}{ccc}[ht!]
\tablecaption{Receiver-sensitivity contribution to the unresolved, minimally drifting EIRP limit \label{tab:sensitivity_range}}
\tablehead{
\colhead{$A_{\rm eff}/T_{\rm sys}$} & \colhead{$S_{\rm min}$} & \colhead{EIRP$_{\rm min}$} \\
\colhead{(m$^{2}$ K$^{-1}$)} & \colhead{(W m$^{-2}$)} & \colhead{(W)}
}
\startdata
1800 & $8.6\times10^{-27}$ & $1.60\times10^{10}$ \\
2000 (nominal) & $7.7\times10^{-27}$ & $1.44\times10^{10}$ \\
2200 & $7.0\times10^{-27}$ & $1.31\times10^{10}$ \\
\enddata
\tablecomments{The range is not an epoch-specific calibration; it is a sensitivity bracket based on published FAST L-band 19-beam performance and the 30.56--35.88 deg midpoint zenith angles of this campaign.}
\end{deluxetable}

A second correction is required for high drift rates. The tree de-Doppler algorithm used by \texttt{turboSETI} can lose peak-channel sensitivity when a signal drifts by more than one frequency channel per time integration \citep{2013ApJ...767...94S,2023AJ..Sheikh.bl,2021AJ....161...55M,2024AJ.Tusay}. For our data, the one-pixel drift rate is
\begin{equation}
    \alpha = \frac{\delta\nu}{\delta t} \approx \frac{7.5~{\rm Hz}}{10~{\rm s}} = 0.75~{\rm Hz~s^{-1}} .
\end{equation}
For $|\dot{\nu}|>\alpha$, we conservatively scale the peak-channel retained sensitivity as $\alpha/|\dot{\nu}|$ and the effective EIRP threshold as $|\dot{\nu}|/\alpha$. This correction does not recover power loss and is not a substitute for a frequency-scrunched or injection-recovery reanalysis; rather, it quantifies the sensitivity degradation of the analysis as performed. The resulting drift-dependent limits are shown in Table~\ref{tab:drift_sensitivity}.

\begin{deluxetable}{cccc}[ht!]
\tablecaption{Drift-rate-dependent peak-channel sensitivity correction for the analysis as performed \label{tab:drift_sensitivity}}
\tablehead{
\colhead{$|\dot{\nu}|$} & \colhead{Retained peak sensitivity} & \colhead{EIRP multiplier} & \colhead{Effective EIRP$_{\rm min}$} \\
\colhead{(Hz s$^{-1}$)} & \colhead{} & \colhead{} & \colhead{(W)}
}
\startdata
0.00 & 1.000 & 1.00 & $1.44\times10^{10}$ \\
0.75 & 1.000 & 1.00 & $1.44\times10^{10}$ \\
1.00 & 0.750 & 1.33 & $1.92\times10^{10}$ \\
2.00 & 0.375 & 2.67 & $3.84\times10^{10}$ \\
5.00 & 0.150 & 6.67 & $9.60\times10^{10}$ \\
10.00 & 0.075 & 13.33 & $1.92\times10^{11}$ \\
20.00 & 0.0375 & 26.67 & $3.84\times10^{11}$ \\
\enddata
\tablecomments{Values use the nominal $1.44\times10^{10}$~W unresolved, minimally drifting limit. The high-drift entries should be interpreted as sensitivity accounting for this search, not as reprocessed high-drift recovery limits.}
\end{deluxetable}

Thus, the headline $1.44\times10^{10}$~W value represents only the nominal unresolved, low-drift limit. The full sensitivity envelope depends on receiver sensitivity, drift rate, and apparent signal bandwidth. We do not assign planet-specific propagation-corrected EIRP limits in this work. The broadening of an intrinsically narrowband signal by the TRAPPIST-1 exoplanetary interplanetary medium (Exo-IPM) depends on line-of-sight impact parameter, orbital phase and geometry, observing frequency, stellar-wind turbulence, and possible coronal mass ejection (CME) conditions, none of which are constrained for our observing epochs. We therefore treat propagation broadening as a linewidth-dependent limitation on the unresolved narrowband sensitivity, rather than as a planet-by-planet correction. Receiver-sensitivity and drift-rate contributions are summarized in Tables~\ref{tab:sensitivity_range} and \ref{tab:drift_sensitivity}, while linewidth-dependent peak-channel sensitivity penalties for assumed apparent widths are summarized in Table~\ref{tab:broadening_penalty}.

\begin{deluxetable}{ccc}[ht!]
\tablecaption{First-order peak-channel sensitivity penalties for assumed apparent linewidths \label{tab:broadening_penalty}}
\tablehead{
\colhead{Apparent linewidth $W_{\rm app}$} & \colhead{Peak-channel multiplier} & \colhead{Effective EIRP threshold} \\
\colhead{(Hz)} & \colhead{} & \colhead{(W)}
}
\startdata
7.5 & 1.00 & $1.44\times10^{10}$ \\
10 & 1.33 & $1.92\times10^{10}$ \\
30 & 4.00 & $5.76\times10^{10}$ \\
100 & 13.33 & $1.92\times10^{11}$ \\
1000 & 133.33 & $1.92\times10^{12}$ \\
\enddata
\tablecomments{The multiplier is a first-order peak-channel dilution factor, $\max(1,W_{\rm app}/7.5~{\rm Hz})$, for the 7.5~Hz channelized narrowband search used here. $W_{\rm app}$ is an assumed effective apparent linewidth. These values are illustrative sensitivity penalties for fixed linewidths; they are not planet-specific Exo-IPM corrections. A physical Exo-IPM correction would require the line-of-sight impact parameter, orbital phase and geometry, observing frequency, stellar-wind turbulence, and possible coronal mass ejection (CME) conditions. An optimized matched-bandwidth analysis would have a weaker $\sqrt{W_{\rm app}/7.5~{\rm Hz}}$ scaling, but that was not the analysis performed here.}
\end{deluxetable}

\section{Discussion}
\label{sec:discussion}

In this work, we carried out a targeted unresolved narrowband search of the TRAPPIST-1 system with five separate pointings (5 $\times$ 20 min; see Section~\ref{sec:obs}). No technosignature candidates survived our selection process within the searched parameter space (1.05--1.45\,GHz, drift rates $|\dot{\nu}|\le 20\,$Hz\,s$^{-1}$, S/N threshold = 10). From these nondetections we derive an observational sensitivity envelope for transmitters located at the distance of TRAPPIST-1. Under the nominal assumptions described in Section~\ref{sec:results}, this corresponds to $\mathrm{EIRP}_{\min}\approx1.44\times10^{10}\,$W for unresolved, minimally drifting narrowband signals, but the effective limit is less stringent for high drift rates, lower receiver sensitivity, or propagation-broadened/resolved signals.

%It is important to emphasize the temporal scope of our constraints. Because we do not possess contemporaneous multi-wavelength (optical/UV/X-ray) monitoring that would unambiguously identify whether each of the five pointings fell during a stellar flare or during a quiescent interval, we cannot claim that our limits apply specifically to a “quiescent (non-flaring) state” of TRAPPIST-1. More precisely, our upper limits are strictly valid for the {\em actual observation windows} sampled by this campaign (the five 20-minute exposures); they should not be extrapolated to make general statements about the system at all epochs or specifically during flare- or post-flare intervals. A statement that implies constraint over a quiescent epoch would therefore be too strong given the available ancillary data.\footnote{A similar caveat about activity-state dependence of detectability has been noted in prior SETI and stellar-activity literature (see e.g. \citealt {Benford2010, Vida2017, Garraffo2017}).}

The narrowband radiometer-equation limits in Section~\ref{sec:results} do not apply unchanged to resolved, broadband, or strongly modulated signals. If the same total transmitter power is spread over many frequency channels, a search optimized for a single unresolved channel will have reduced peak S/N. With an optimal matched-bandwidth search, the total-power threshold scales approximately as $\sqrt{W_{\rm app}/\delta\nu}$, where $W_{\rm app}$ is the apparent signal bandwidth; with a peak-channel narrowband search like the one used here, the effective threshold can scale more conservatively as $W_{\rm app}/\delta\nu$. Therefore, broadband or spectrally resolved transmitters could be missed, fragmented into multiple hits, or rejected as RFI-like morphology by a pipeline optimized for persistent Hz-scale tracks.

Propagation through the TRAPPIST-1 exoplanetary interplanetary medium is another important limitation. \citet{2026ApJ...999..210G} show that turbulence in exoplanetary interplanetary media (Exo-IPM), especially around active M-dwarf systems, can redistribute an intrinsically narrowband signal into broader spectral wings and suppress the peak S/N targeted by standard narrowband pipelines. The exact broadening for TRAPPIST-1 depends on stellar-wind properties, CME occurrence, observing frequency, orbital phase, and line-of-sight impact parameter and geometry, none of which are constrained for our observing epochs. We therefore do not claim a planet-specific Exo-IPM correction. Instead, Table~\ref{tab:broadening_penalty} gives first-order peak-channel sensitivity penalties for assumed apparent linewidths. In a quiescent optimistic case where the apparent width remains within one 7.5~Hz channel, the nominal limit is unchanged. If the apparent width reaches 100--1000~Hz, the peak-channel EIRP threshold would increase by factors of 13--133 for the analysis as performed. These values are sensitivity penalties for assumed linewidths, not physical Exo-IPM-corrected limits for individual TRAPPIST-1 planets.

Several additional classes of signals remain poorly constrained by our observing strategy. Sparse temporal sampling (short-duration pointings separated in time) and the finite on-source time reduce our completeness for transmitters that (i) operate with very low duty cycle ($\tau \ll 20\,$min), (ii) emit extremely narrow time-domain pulses, (iii) hop rapidly in frequency rather than remaining as a persistent drifting narrowband track, or (iv) use highly directive beams that rarely intersect Earth. Such signals would require either very high instantaneous EIRP to be detected in our sparse sampling, or dedicated strategies optimized for transients (e.g. higher time resolution recording, real-time triggering), frequency-agile emission, and/or long-duration continuous monitoring to raise detection probability \citep{2018AJ..Wright,2010AsBio.Benford}.

Stellar activity in M-dwarf systems like TRAPPIST-1 has twofold implications for technosignature searches. On the one hand, frequent flaring can increase radio background/contamination, produce plasma conditions that broaden narrowband signals, and complicate discrimination between natural bursts and candidate artificial signals; on the other hand, astrophysical flares could plausibly serve as coordination or trigger events for intentional transmissions (a Schelling-point-like strategy), motivating targeted ``flare-triggered'' observing programs \citep{schelling1980strategy,2017ApJ..Vida,2026ApJ...999..210G}. Our present campaign, lacking contemporaneous flare diagnostics, does not directly test flare-triggered hypotheses and therefore leaves this observationally motivated parameter space open.

Given its proximity and seven Earth-sized planets, several of which reside in the habitable zone, TRAPPIST-1 remains one of the most compelling nearby systems for SETI investigations and a valuable benchmark for comparative studies. Looking ahead, we will continue to observe TRAPPIST-1 and expand our search to include additional classes of technosignatures (for example, periodic signals; \citealt{2023AJ....165..255S}). Future analyses should include injection-recovery tests across drift rate and apparent bandwidth, frequency-scrunched or otherwise high-drift-optimized searches, and contemporaneous stellar-activity monitoring to better account for Exo-IPM propagation. We will also apply the same survey strategy to a larger sample of nearby stars and exoplanetary systems with FAST to build statistical coverage across targets.

\section
{Conclusions}
\label{sec: Conclusions}

We conducted an unresolved narrowband SETI search toward the TRAPPIST-1 system using five FAST L-band pointings (5 $\times$ 20\,min), covering 1.05--1.45\,GHz with drift rates $|\dot{\nu}|\le 20\,$Hz\,s$^{-1}$ and a detection threshold of S/N = 10. No credible technosignature candidates attributable to the TRAPPIST-1 system were identified in the searched parameter space. From these nondetections we derive a nominal minimum detectable equivalent isotropic radiated power of $\mathrm{EIRP}_{\min}\approx1.44\times10^{10}\,$W for unresolved, minimally drifting, non-broadened narrowband signals at the distance of TRAPPIST-1. This value should not be interpreted as a uniform limit across the full drift-rate range: after accounting for tree de-Doppler power smearing, the effective limit reaches $\approx3.84\times10^{11}\,$W at $|\dot{\nu}|=20\,\mathrm{Hz\,s^{-1}}$.

Despite these constraints, several classes of signals remain poorly probed by our current strategy, including transmitters with very low duty cycle, extremely short-duration pulses, broadband or propagation-broadened emission, rapid frequency agility, or highly directive beams. A physical Exo-IPM correction would require epoch-specific propagation modeling and is not attempted here. Future work will continue to observe TRAPPIST-1 and expand the search to other nearby stars and exoplanetary systems with FAST, including additional classes of technosignatures such as periodic signals \citep{2023AJ....165..255S}. A revised analysis with injection recovery, high-drift mitigation, and propagation-aware sensitivity accounting would further improve the quantitative interpretation of FAST nondetections in active M-dwarf planetary systems.

\vspace{12 pt}
We sincerely appreciate the insightful and useful comments of the referee and editor, which helped us improve our manuscript. This research was supported by the funding from the National SKA Program of China under Grant No. 2025SKA0120104, Shandong Provincial Natural Science Foundation (ZR2024QA180), Scientific Research Fund of Dezhou University 4022504019, National Key R\&D Program of China, No. 2024YFA1611804, and the China Manned Space Program with grant No. CMS-CSST-2025-A01.
\bibliography{text}

@ARTICLE{1959Natur.184..844C,
       author = {{Cocconi}, Giuseppe and {Morrison}, Philip},
        title = "{Searching for Interstellar Communications}",
      journal = {\nat},
         year = 1959,
        month = sep,
       volume = {184},
       number = {4690},
        pages = {844-846},
          doi = {10.1038/184844a0},
       adsurl = {https://ui.adsabs.harvard.edu/abs/1959Natur.184..844C}
}

@ARTICLE{2020RAA....20...64J,
       author = {{Jiang}, Peng and {Tang}, Ning-Yu and {Hou}, Li-Gang and {Liu}, Meng-Ting and {Kr{\v{c}}o}, Marko and {Qian}, Lei and {Sun}, Jing-Hai and {Ching}, Tao-Chung and {Liu}, Bin and {Duan}, Yan and {Yue}, You-Ling and {Gan}, Heng-Qian and {Yao}, Rui and {Li}, Hui and {Pan}, Gao-Feng and {Yu}, Dong-Jun and {Liu}, Hong-Fei and {Li}, Di and {Peng}, Bo and {Yan}, Jun and {FAST Collaboration}},
        title = "{The fundamental performance of FAST with 19-beam receiver at L band}",
      journal = {Research in Astronomy and Astrophysics},
         year = 2020,
        month = may,
       volume = {20},
       number = {5},
          eid = {064},
        pages = {064},
          doi = {10.1088/1674-4527/20/5/64},
archivePrefix = {arXiv},
       eprint = {2002.01786},
 primaryClass = {astro-ph.IM},
       adsurl = {https://ui.adsabs.harvard.edu/abs/2020RAA....20...64J}
}

@ARTICLE{2011IJMPD..20..989N,
       author = {{Nan}, Rendong and {Li}, Di and {Jin}, Chengjin and {Wang}, Qiming and {Zhu}, Lichun and {Zhu}, Wenbai and {Zhang}, Haiyan and {Yue}, Youling and {Qian}, Lei},
        title = "{The Five-Hundred Aperture Spherical Radio Telescope (fast) Project}",
      journal = {International Journal of Modern Physics D},
         year = 2011,
        month = jan,
       volume = {20},
       number = {6},
        pages = {989-1024},
          doi = {10.1142/S0218271811019335},
archivePrefix = {arXiv},
       eprint = {1105.3794},
 primaryClass = {astro-ph.IM},
       adsurl = {https://ui.adsabs.harvard.edu/abs/2011IJMPD..20..989N}
}

@ARTICLE{2021A&A...649A...1G,
       author = {{Gaia Collaboration} and {Brown}, A.~G.~A. and {Vallenari}, A. and {Prusti}, T. and {de Bruijne}, J.~H.~J. and {Babusiaux}, C. and {Biermann}, M. and {Creevey}, O.~L. and {Evans}, D.~W. and {Eyer}, L. and et al.},
        title = "{Gaia Early Data Release 3. Summary of the contents and survey properties}",
      journal = {\aap},
         year = 2021,
        month = may,
       volume = {649},
          eid = {A1},
        pages = {A1},
          doi = {10.1051/0004-6361/202039657},
archivePrefix = {arXiv},
       eprint = {2012.01533},
 primaryClass = {astro-ph.GA},
       adsurl = {https://ui.adsabs.harvard.edu/abs/2021A&A...649A...1G}
}

@ARTICLE{2020ApJ...891..174Z,
       author = {{Zhang}, Zhi-Song and {Werthimer}, Dan and {Zhang}, Tong-Jie and {Cobb}, Jeff and {Korpela}, Eric and {Anderson}, David and {Gajjar}, Vishal and {Lee}, Ryan and {Li}, Shi-Yu and {Pei}, Xin and {Zhang}, Xin-Xin and {Huang}, Shi-Jie and {Wang}, Pei and {Zhu}, Yan and {Duan}, Ran and {Zhang}, Hai-Yan and {Jin}, Cheng-jin and {Zhu}, Li-Chun and {Li}, Di},
        title = "{First SETI Observations with China's Five-hundred-meter Aperture Spherical Radio Telescope (FAST)}",
      journal = {\apj},
         year = 2020,
        month = mar,
       volume = {891},
       number = {2},
          eid = {174},
        pages = {174},
          doi = {10.3847/1538-4357/ab7376},
archivePrefix = {arXiv},
       eprint = {2002.02130},
 primaryClass = {astro-ph.IM},
       adsurl = {https://ui.adsabs.harvard.edu/abs/2020ApJ...891..174Z}
}

@ARTICLE{2016RaSc...51.1060L,
       author = {{Li}, Di and {Pan}, Zhichen},
        title = "{The Five-hundred-meter Aperture Spherical Radio Telescope Project}",
      journal = {Radio Science},
         year = 2016,
        month = jul,
       volume = {51},
        pages = {1060-1064},
          doi = {10.1002/2015RS005877},
archivePrefix = {arXiv},
       eprint = {1612.09372},
 primaryClass = {astro-ph.IM},
       adsurl = {https://ui.adsabs.harvard.edu/abs/2016RaSc...51.1060L}
}

@ARTICLE{2001ARA&A..39..511T,
       author = {{Tarter}, Jill},
        title = "{The Search for Extraterrestrial Intelligence (SETI)}",
      journal = {\araa},
         year = 2001,
        month = jan,
       volume = {39},
        pages = {511-548},
          doi = {10.1146/annurev.astro.39.1.511},
       adsurl = {https://ui.adsabs.harvard.edu/abs/2001ARA&A..39..511T}
}

@ARTICLE{2013ApJ...767...94S,
       author = {{Siemion}, Andrew P.~V. and {Demorest}, Paul and {Korpela}, Eric and {Maddalena}, Ron J. and {Werthimer}, Dan and {Cobb}, Jeff and {Howard}, Andrew W. and {Langston}, Glen and {Lebofsky}, Matt and {Marcy}, Geoffrey W. and {Tarter}, Jill},
        title = "{A 1.1-1.9 GHz SETI Survey of the Kepler Field. I. A Search for Narrow-band Emission from Select Targets}",
      journal = {\apj},
         year = 2013,
        month = apr,
       volume = {767},
       number = {1},
          eid = {94},
        pages = {94},
          doi = {10.1088/0004-637X/767/1/94},
archivePrefix = {arXiv},
       eprint = {1302.0845},
 primaryClass = {astro-ph.GA},
       adsurl = {https://ui.adsabs.harvard.edu/abs/2013ApJ...767...94S}
}

@ARTICLE{2019JOSS....4.1554P,
       author = {{Price}, Danny and {Enriquez}, J. and {Chen}, Yuhong and {Siebert}, Mark},
        title = "{Blimpy: Breakthrough Listen I/O Methods for Python}",
      journal = {The Journal of Open Source Software},
         year = 2019,
        month = oct,
       volume = {4},
       number = {42},
          eid = {1554},
        pages = {1554},
          doi = {10.21105/joss.01554},
       adsurl = {https://ui.adsabs.harvard.edu/abs/2019JOSS....4.1554P}
}

@MISC{2019ascl.soft06006E,
       author = {{Enriquez}, Emilio and {Price}, Danny},
        title = "{turboSETI: Python-based SETI search algorithm}",
         year = 2019,
        month = jun,
          eid = {ascl:1906.006},
        pages = {ascl:1906.006},
archivePrefix = {ascl},
       eprint = {1906.006},
       adsurl = {https://ui.adsabs.harvard.edu/abs/2019ascl.soft06006E}
}

@ARTICLE{2021RAA....21...18W,
       author = {{Wang}, Yu and {Zhang}, Hai-Yan and {Hu}, Hao and {Huang}, Shi-Jie and {Zhu}, Wei-Wei and {Zhi}, Guo-Ping and {Zhang}, Tao and {Fan}, Zhan-Chun and {Yang}, Li},
        title = "{Satellite RFI mitigation on FAST}",
      journal = {Research in Astronomy and Astrophysics},
         year = 2021,
        month = jan,
       volume = {21},
       number = {1},
          eid = {018},
        pages = {018},
          doi = {10.1088/1674-4527/21/1/18},
       adsurl = {https://ui.adsabs.harvard.edu/abs/2021RAA....21...18W}
}

@ARTICLE{2021AJ....162...33G,
       author = {{Gajjar}, Vishal and {Perez}, Karen I. and {Siemion}, Andrew P.~V. and {Foster}, Griffin and {Brzycki}, Bryan and {Chatterjee}, Shami and {Chen}, Yuhong and {Cordes}, James M. and {Croft}, Steve and {Czech}, Daniel and {DeBoer}, David and {DeMarines}, Julia and {Drew}, Jamie and {Gowanlock}, Michael and {Isaacson}, Howard and {Lacki}, Brian C. and {Lebofsky}, Matt and {MacMahon}, David H.~E. and {Morrison}, Ian S. and {Ng}, Cherry and {de Pater}, Imke and {Price}, Danny C. and {Sheikh}, Sofia Z. and {Suresh}, Akshay and {Webb}, Claire and {Pete Worden}, S.},
        title = "{The Breakthrough Listen Search For Intelligent Life Near the Galactic Center. I.}",
      journal = {\aj},
         year = 2021,
        month = jul,
       volume = {162},
       number = {1},
          eid = {33},
        pages = {33},
          doi = {10.3847/1538-3881/abfd36},
archivePrefix = {arXiv},
       eprint = {2104.14148},
 primaryClass = {astro-ph.HE},
       adsurl = {https://ui.adsabs.harvard.edu/abs/2021AJ....162...33G}
}

@ARTICLE{2017ApJ...849..104E,
       author = {{Enriquez}, J. Emilio and {Siemion}, Andrew and {Foster}, Griffin and {Gajjar}, Vishal and {Hellbourg}, Greg and {Hickish}, Jack and {Isaacson}, Howard and {Price}, Danny C. and {Croft}, Steve and {DeBoer}, David and {Lebofsky}, Matt and {MacMahon}, David H.~E. and {Werthimer}, Dan},
        title = "{The Breakthrough Listen Search for Intelligent Life: 1.1-1.9 GHz Observations of 692 Nearby Stars}",
      journal = {\apj},
         year = 2017,
        month = nov,
       volume = {849},
       number = {2},
          eid = {104},
        pages = {104},
          doi = {10.3847/1538-4357/aa8d1b},
archivePrefix = {arXiv},
       eprint = {1709.03491},
 primaryClass = {astro-ph.EP},
       adsurl = {https://ui.adsabs.harvard.edu/abs/2017ApJ...849..104E}
}

@ARTICLE{2020AJ....159...86P,
       author = {{Price}, Danny C. and {Enriquez}, J. Emilio and {Brzycki}, Bryan and {Croft}, Steve and {Czech}, Daniel and {DeBoer}, David and {DeMarines}, Julia and {Foster}, Griffin and {Gajjar}, Vishal and {Gizani}, Nectaria and {Hellbourg}, Greg and {Isaacson}, Howard and {Lacki}, Brian and {Lebofsky}, Matt and {MacMahon}, David H.~E. and {Pater}, Imke de and {Siemion}, Andrew P.~V. and {Werthimer}, Dan and {Green}, James A. and {Kaczmarek}, Jane F. and {Maddalena}, Ronald J. and {Mader}, Stacy and {Drew}, Jamie and {Worden}, S. Pete},
        title = "{The Breakthrough Listen Search for Intelligent Life: Observations of 1327 Nearby Stars Over 1.10-3.45 GHz}",
      journal = {\aj},
         year = 2020,
        month = mar,
       volume = {159},
       number = {3},
          eid = {86},
        pages = {86},
          doi = {10.3847/1538-3881/ab65f1},
archivePrefix = {arXiv},
       eprint = {1906.07750},
 primaryClass = {astro-ph.EP},
       adsurl = {https://ui.adsabs.harvard.edu/abs/2020AJ....159...86P}
}

@ARTICLE{2022AJ....164..160T,
       author = {{Tao}, Zhen-Zhao and {Zhao}, Hai-Chen and {Zhang}, Tong-Jie and {Gajjar}, Vishal and {Zhu}, Yan and {Yue}, You-Ling and {Zhang}, Hai-Yan and {Liu}, Wen-Fei and {Li}, Shi-Yu and {Zhang}, Jian-Chen and {Liu}, Cong and {Wang}, Hong-Feng and {Duan}, Ran and {Qian}, Lei and {Jin}, Cheng-Jin and {Li}, Di and {Siemion}, Andrew and {Jiang}, Peng and {Werthimer}, Dan and {Cobb}, Jeff and {Korpela}, Eric and {Anderson}, David P.},
        title = "{Sensitive Multibeam Targeted SETI Observations toward 33 Exoplanet Systems with FAST}",
      journal = {\aj},
         year = 2022,
        month = oct,
       volume = {164},
       number = {4},
          eid = {160},
        pages = {160},
          doi = {10.3847/1538-3881/ac8bd5},
archivePrefix = {arXiv},
       eprint = {2208.02421},
 primaryClass = {astro-ph.EP},
       adsurl = {https://ui.adsabs.harvard.edu/abs/2022AJ....164..160T}
}

@ARTICLE{2023AJ....165..132L,
       author = {{Luan}, Xiao-Hang and {Tao}, Zhen-Zhao and {Zhao}, Hai-Chen and {Huang}, Bo-Lun and {Li}, Shi-Yu and {Liu}, Cong and {Wang}, Hong-Feng and {Liu}, Wen-Fei and {Zhang}, Tong-Jie and {Gajjar}, Vishal and {Werthimer}, Dan},
        title = "{Multibeam Blind Search of Targeted SETI Observations toward 33 Exoplanet Systems with FAST}",
      journal = {\aj},
         year = 2023,
        month = mar,
       volume = {165},
       number = {3},
          eid = {132},
        pages = {132},
          doi = {10.3847/1538-3881/acb706},
archivePrefix = {arXiv},
       eprint = {2301.10890},
 primaryClass = {astro-ph.EP},
       adsurl = {https://ui.adsabs.harvard.edu/abs/2023AJ....165..132L}
}

@ARTICLE{2023AJ....165..255S,
       author = {{Suresh}, Akshay and {Gajjar}, Vishal and {Nagarajan}, Pranav and {Sheikh}, Sofia Z. and {Siemion}, Andrew P.~V. and {Lebofsky}, Matt and {MacMahon}, David H.~E. and {Price}, Danny C. and {Croft}, Steve},
        title = "{A 4-8 GHz Galactic Center Search for Periodic Technosignatures}",
      journal = {\aj},
         year = 2023,
        month = jun,
       volume = {165},
       number = {6},
          eid = {255},
        pages = {255},
          doi = {10.3847/1538-3881/acccf0},
archivePrefix = {arXiv},
       eprint = {2305.18527},
 primaryClass = {astro-ph.IM},
       adsurl = {https://ui.adsabs.harvard.edu/abs/2023AJ....165..255S}
}

@ARTICLE{1997ApJ...487..782C,
       author = {{Cordes}, James M. and {Lazio}, Joseph W. and {Sagan}, Carl},
        title = "{Scintillation-induced Intermittency in SETI}",
      journal = {\apj},
         year = 1997,
        month = oct,
       volume = {487},
       number = {2},
        pages = {782-808},
          doi = {10.1086/304620},
archivePrefix = {arXiv},
       eprint = {astro-ph/9707039},
 primaryClass = {astro-ph},
       adsurl = {https://ui.adsabs.harvard.edu/abs/1997ApJ...487..782C}
}

@ARTICLE{1981A&AS...44..363W,
       author = {{Wells}, D.~C. and {Greisen}, E.~W. and {Harten}, R.~H.},
        title = "{FITS - a Flexible Image Transport System}",
      journal = {\aaps},
         year = 1981,
        month = jun,
       volume = {44},
        pages = {363},
       adsurl = {https://ui.adsabs.harvard.edu/abs/1981A&AS...44..363W}
}

@ARTICLE{2025AJ..luan,
       author = {{Luan}, Xiao-Hang and {Huang}, Bo-Lun and {Tao}, Zhen-Zhao and {Cui}, Yan and {Zhang}, Tong-Jie and {Wang}, Pei},
        title = "{Multibeam SETI Observations Toward Nearby M Dwarfs with FAST}",
      journal = {\aj},
         year = 2025,
        month = apr,
       volume = {169},
       number = {4},
          eid = {217},
        pages = {217},
          doi = {10.3847/1538-3881/adbaef},
archivePrefix = {arXiv},
       eprint = {2502.20419},
 primaryClass = {astro-ph.IM},
       adsurl = {https://ui.adsabs.harvard.edu/abs/2025AJ....169..217L}
}

@ARTICLE{2023AJ..huang,
       author = {{Huang}, Bo-Lun and {Tao}, Zhen-Zhao and {Zhang}, Tong-Jie},
        title = "{A Solution to Continuous RFI in Narrowband Radio SETI with FAST: The MultiBeam Point-source Scanning Strategy}",
      journal = {\aj},
         year = 2023,
        month = dec,
       volume = {166},
       number = {6},
          eid = {245},
        pages = {245},
          doi = {10.3847/1538-3881/ad06b1},
archivePrefix = {arXiv},
       eprint = {2307.11368},
 primaryClass = {astro-ph.IM},
       adsurl = {https://ui.adsabs.harvard.edu/abs/2023AJ....166..245H}
}

@ARTICLE{2023AJ..tao,
       author = {{Tao}, Zhen-Zhao and {Huang}, Bo-Lun and {Luan}, Xiao-Hang and {Li}, Jian-Kang and {Zhao}, Hai-Chen and {Wang}, Hong-Feng and {Zhang}, Tong-Jie},
        title = "{The Most Sensitive SETI Observations Toward Barnard's Star with FAST}",
      journal = {\aj},
         year = 2023,
        month = nov,
       volume = {166},
       number = {5},
          eid = {190},
        pages = {190},
          doi = {10.3847/1538-3881/acfc1e},
archivePrefix = {arXiv},
       eprint = {2309.15377},
 primaryClass = {astro-ph.IM},
       adsurl = {https://ui.adsabs.harvard.edu/abs/2023AJ....166..190T}
}

@ARTICLE{2017Natur.Gillon,
       author = {{Gillon}, Micha{\"e}l and {Triaud}, Amaury H.~M.~J. and {Demory}, Brice-Olivier and {Jehin}, Emmanu{\"e}l and {Agol}, Eric and {Deck}, Katherine M. and {Lederer}, Susan M. and {de Wit}, Julien and {Burdanov}, Artem and {Ingalls}, James G. and {Bolmont}, Emeline and {Leconte}, Jeremy and {Raymond}, Sean N. and {Selsis}, Franck and {Turbet}, Martin and {Barkaoui}, Khalid and {Burgasser}, Adam and {Burleigh}, Matthew R. and {Carey}, Sean J. and {Chaushev}, Aleksander and {Copperwheat}, Chris M. and {Delrez}, Laetitia and {Fernandes}, Catarina S. and {Holdsworth}, Daniel L. and {Kotze}, Enrico J. and {Van Grootel}, Val{\'e}rie and {Almleaky}, Yaseen and {Benkhaldoun}, Zouhair and {Magain}, Pierre and {Queloz}, Didier},
        title = "{Seven temperate terrestrial planets around the nearby ultracool dwarf star TRAPPIST-1}",
      journal = {\nat},
         year = 2017,
        month = feb,
       volume = {542},
       number = {7642},
        pages = {456-460},
          doi = {10.1038/nature21360},
archivePrefix = {arXiv},
       eprint = {1703.01424},
 primaryClass = {astro-ph.EP},
       adsurl = {https://ui.adsabs.harvard.edu/abs/2017Natur.542..456G}
}

@ARTICLE{2024AJ.Tusay,
       author = {{Tusay}, Nick and {Sheikh}, Sofia Z. and {Sneed}, Evan L. and {Farah}, Wael and {Pollak}, Alexander W. and {Cruz}, Luigi F. and {Siemion}, Andrew and {DeBoer}, David R. and {Wright}, Jason T.},
        title = "{A Radio Technosignature Search of TRAPPIST-1 with the Allen Telescope Array}",
      journal = {\aj},
         year = 2024,
        month = dec,
       volume = {168},
       number = {6},
          eid = {283},
        pages = {283},
          doi = {10.3847/1538-3881/ad823c},
archivePrefix = {arXiv},
       eprint = {2409.08313},
 primaryClass = {astro-ph.EP},
       adsurl = {https://ui.adsabs.harvard.edu/abs/2024AJ....168..283T}
}

@INPROCEEDINGS{2000ASPC.Cobb.sh,
       author = {{Cobb}, Jeff and {Lebofsky}, M. and {Werthimer}, D. and {Bowyer}, S. and {Lampton}, M.},
        title = "{SERENDIP IV: Data Acquisition, Reduction, and Analysis}",
    booktitle = {Bioastronomy 99},
         year = 2000,
       editor = {{Lemarchand}, Guillermo and {Meech}, Karen},
       series = {Astronomical Society of the Pacific Conference Series},
       volume = {213},
        month = jan,
        pages = {485},
       adsurl = {https://ui.adsabs.harvard.edu/abs/2000ASPC..213..485C}
}

@ARTICLE{2001CSE..Korpela.sh,
       author = {{Dubois}, Paul F. and {Korpela}, Eric and {Werthimer}, Dan and {Anderson}, David and {Cobb}, Jeff and {Lebofsky}, Matt},
        title = "{Seti@Home{\textemdash}Massively Distributed Computing for Seti}",
      journal = {Computing in Science and Engineering},
         year = 2001,
        month = jan,
       volume = {3},
       number = {1},
        pages = {78-83},
          doi = {10.1109/5992.895191},
       adsurl = {https://ui.adsabs.harvard.edu/abs/2001CSE.....3a..78K}
}

@ARTICLE{2025AJ..Anderson.sh,
       author = {{Anderson}, David P. and {Korpela}, Eric J. and {Werthimer}, Dan and {Cobb}, Jeff and {Allen}, Bruce},
        title = "{SETI@home: Data Analysis and Findings}",
      journal = {\aj},
         year = 2025,
        month = aug,
       volume = {170},
       number = {2},
          eid = {111},
        pages = {111},
          doi = {10.3847/1538-3881/ade5ab},
archivePrefix = {arXiv},
       eprint = {2506.14737},
 primaryClass = {astro-ph.IM},
       adsurl = {https://ui.adsabs.harvard.edu/abs/2025AJ....170..111A}
}

@ARTICLE{2025AJ..Korpela.sh,
       author = {{Korpela}, E.~J. and {Anderson}, D.~P. and {Cobb}, J. and {Lebofsky}, M. and {Liu}, W. and {Werthimer}, D.},
        title = "{SETI@home: Data Acquisition and Front-end Processing}",
      journal = {\aj},
         year = 2025,
        month = aug,
       volume = {170},
       number = {2},
          eid = {112},
        pages = {112},
          doi = {10.3847/1538-3881/ade5a7},
archivePrefix = {arXiv},
       eprint = {2506.14718},
 primaryClass = {astro-ph.IM},
       adsurl = {https://ui.adsabs.harvard.edu/abs/2025AJ....170..112K}
}

@ARTICLE{2023AJ..Sheikh.bl,
       author = {{Sheikh}, Sofia Z. and {Kanodia}, Shubham and {Lubar}, Emily and {Bowman}, William P. and {Ca{\~n}as}, Caleb I. and {Gilbertson}, Christian and {MacDonald}, Mariah G. and {Wright}, Jason and {MacMahon}, David and {Croft}, Steve and {Price}, Danny and {Siemion}, Andrew and {Drew}, Jamie and {Worden}, S. Pete and {Trenholm}, Elizabeth and {The Breakthrough Listen Initiative}},
        title = "{A Green Bank Telescope Search for Narrowband Technosignatures between 1.1 and 1.9 GHz During 12 Kepler Planetary Transits}",
      journal = {\aj},
         year = 2023,
        month = feb,
       volume = {165},
       number = {2},
          eid = {61},
        pages = {61},
          doi = {10.3847/1538-3881/aca907},
archivePrefix = {arXiv},
       eprint = {2212.05137},
 primaryClass = {astro-ph.EP},
       adsurl = {https://ui.adsabs.harvard.edu/abs/2023AJ....165...61S}
}

@ARTICLE{2023NatAs.bl,
       author = {{Ma}, Peter Xiangyuan and {Ng}, Cherry and {Rizk}, Leandro and {Croft}, Steve and {Siemion}, Andrew P.~V. and {Brzycki}, Bryan and {Czech}, Daniel and {Drew}, Jamie and {Gajjar}, Vishal and {Hoang}, John and {Isaacson}, Howard and {Lebofsky}, Matt and {MacMahon}, David H.~E. and {de Pater}, Imke and {Price}, Danny C. and {Sheikh}, Sofia Z. and {Worden}, S. Pete},
        title = "{A deep-learning search for technosignatures from 820 nearby stars}",
      journal = {Nature Astronomy},
         year = 2023,
        month = apr,
       volume = {7},
        pages = {492-502},
          doi = {10.1038/s41550-022-01872-z},
archivePrefix = {arXiv},
       eprint = {2301.12670},
 primaryClass = {astro-ph.IM},
       adsurl = {https://ui.adsabs.harvard.edu/abs/2023NatAs...7..492M}
}

@ARTICLE{2025AJ..Pardo.bl,
       author = {{Pardo}, Snir and {Poznanski}, Dovi and {Croft}, Steve and {Siemion}, Andrew P.~V. and {Lebofsky}, Matthew},
        title = "{Using Anomaly Detection to Search for Technosignatures in Breakthrough Listen Observations}",
      journal = {\aj},
         year = 2025,
        month = jul,
       volume = {170},
       number = {1},
          eid = {12},
        pages = {12},
          doi = {10.3847/1538-3881/add52b},
archivePrefix = {arXiv},
       eprint = {2505.03927},
 primaryClass = {astro-ph.IM},
       adsurl = {https://ui.adsabs.harvard.edu/abs/2025AJ....170...12P}
}

@ARTICLE{2018AJ..Wright,
       author = {{Wright}, Jason T. and {Kanodia}, Shubham and {Lubar}, Emily},
        title = "{How Much SETI Has Been Done? Finding Needles in the n-dimensional Cosmic Haystack}",
      journal = {\aj},
         year = 2018,
        month = dec,
       volume = {156},
       number = {6},
          eid = {260},
        pages = {260},
          doi = {10.3847/1538-3881/aae099},
archivePrefix = {arXiv},
       eprint = {1809.07252},
 primaryClass = {astro-ph.IM},
       adsurl = {https://ui.adsabs.harvard.edu/abs/2018AJ....156..260W}
}

@ARTICLE{2017ApJ..Vida,
       author = {{Vida}, K. and {K{\H{o}}v{\'a}ri}, Zs. and {P{\'a}l}, A. and {Ol{\'a}h}, K. and {Kriskovics}, L.},
        title = "{Frequent Flaring in the TRAPPIST-1 System{\textemdash}Unsuited for Life?}",
      journal = {\apj},
         year = 2017,
        month = jun,
       volume = {841},
       number = {2},
          eid = {124},
        pages = {124},
          doi = {10.3847/1538-4357/aa6f05},
archivePrefix = {arXiv},
       eprint = {1703.10130},
 primaryClass = {astro-ph.SR},
       adsurl = {https://ui.adsabs.harvard.edu/abs/2017ApJ...841..124V}
}

@ARTICLE{2010AsBio.Benford,
       author = {{Benford}, James and {Benford}, Gregory and {Benford}, Dominic},
        title = "{Messaging with Cost-Optimized Interstellar Beacons}",
      journal = {Astrobiology},
         year = 2010,
        month = jun,
       volume = {10},
       number = {5},
        pages = {475-490},
          doi = {10.1089/ast.2009.0393},
       adsurl = {https://ui.adsabs.harvard.edu/abs/2010AsBio..10..475B}
}

@book{schelling1980strategy,
  title={The Strategy of Conflict: with a new Preface by the Author},
  author={Schelling, Thomas C},
  year={1980},
  publisher={Harvard university press}
}

@ARTICLE{2020SSRv.Turbet,
       author = {{Turbet}, Martin and {Bolmont}, Emeline and {Bourrier}, Vincent and {Demory}, Brice-Olivier and {Leconte}, J{\'e}r{\'e}my and {Owen}, James and {Wolf}, Eric T.},
        title = "{A Review of Possible Planetary Atmospheres in the TRAPPIST-1 System}",
      journal = {\ssr},
         year = 2020,
        month = jul,
       volume = {216},
       number = {5},
          eid = {100},
        pages = {100},
          doi = {10.1007/s11214-020-00719-1},
archivePrefix = {arXiv},
       eprint = {2007.03334},
 primaryClass = {astro-ph.EP},
       adsurl = {https://ui.adsabs.harvard.edu/abs/2020SSRv..216..100T}
}

@ARTICLE{2021PSJ..Agol,
       author = {{Agol}, Eric and {Dorn}, Caroline and {Grimm}, Simon L. and {Turbet}, Martin and {Ducrot}, Elsa and {Delrez}, Laetitia and {Gillon}, Micha{\"e}l and {Demory}, Brice-Olivier and {Burdanov}, Artem and {Barkaoui}, Khalid and {Benkhaldoun}, Zouhair and {Bolmont}, Emeline and {Burgasser}, Adam and {Carey}, Sean and {de Wit}, Julien and {Fabrycky}, Daniel and {Foreman-Mackey}, Daniel and {Haldemann}, Jonas and {Hernandez}, David M. and {Ingalls}, James and {Jehin}, Emmanuel and {Langford}, Zachary and {Leconte}, J{\'e}r{\'e}my and {Lederer}, Susan M. and {Luger}, Rodrigo and {Malhotra}, Renu and {Meadows}, Victoria S. and {Morris}, Brett M. and {Pozuelos}, Francisco J. and {Queloz}, Didier and {Raymond}, Sean N. and {Selsis}, Franck and {Sestovic}, Marko and {Triaud}, Amaury H.~M.~J. and {Van Grootel}, Valerie},
        title = "{Refining the Transit-timing and Photometric Analysis of TRAPPIST-1: Masses, Radii, Densities, Dynamics, and Ephemerides}",
      journal = {\psj},
         year = 2021,
        month = feb,
       volume = {2},
       number = {1},
          eid = {1},
        pages = {1},
          doi = {10.3847/PSJ/abd022},
archivePrefix = {arXiv},
       eprint = {2010.01074},
 primaryClass = {astro-ph.EP},
       adsurl = {https://ui.adsabs.harvard.edu/abs/2021PSJ.....2....1A}
}

@ARTICLE{2021AJ....161...55M,
       author = {{Margot}, Jean-Luc and {Pinchuk}, Pavlo and {Geil}, Robert and {Alexander}, Stephen and {Arora}, Sparsh and {Biswas}, Swagata and {Cebreros}, Jose and {Desai}, Sanjana Prabhu and {Duclos}, Benjamin and {Dunne}, Riley and {Lin Fu}, Kristy Kwan and {Goel}, Shashwat and {Gonzales}, Julia and {Gonzalez}, Alexander and {Jain}, Rishabh and {Lam}, Adrian and {Lewis}, Briley and {Lewis}, Rebecca and {Li}, Grace and {MacDougall}, Mason and {Makarem}, Christopher and {Manan}, Ivan and {Molina}, Eden and {Nagib}, Caroline and {Neville}, Kyle and {O'Toole}, Connor and {Rockwell}, Valerie and {Rokushima}, Yoichiro and {Romanek}, Griffin and {Schmidgall}, Carlyn and {Seth}, Samar and {Shah}, Rehan and {Shimane}, Yuri and {Singhal}, Myank and {Tokadjian}, Armen and {Villafana}, Lizvette and {Wang}, Zhixian and {Yun}, In and {Zhu}, Lujia and {Lynch}, Ryan S.},
        title = "{A Search for Technosignatures around 11,680 Stars with the Green Bank Telescope at 1.15-1.73 GHz}",
      journal = {\aj},
         year = 2021,
        month = feb,
       volume = {161},
       number = {2},
          eid = {55},
        pages = {55},
          doi = {10.3847/1538-3881/abcc77},
       adsurl = {https://ui.adsabs.harvard.edu/abs/2021AJ....161...55M}
}

@ARTICLE{2023AJ....166..182L,
       author = {{Li}, Megan G. and {Sheikh}, Sofia Z. and {Gilbertson}, Christian and {He}, Matthias Y. and {Isaacson}, Howard and {Croft}, Steve and {Sneed}, Evan L.},
        title = "{Developing a Drift Rate Distribution for Technosignature Searches of Exoplanets}",
      journal = {\aj},
         year = 2023,
        month = nov,
       volume = {166},
       number = {5},
          eid = {182},
        pages = {182},
          doi = {10.3847/1538-3881/acf83d},
archivePrefix = {arXiv},
       eprint = {2311.01427},
 primaryClass = {astro-ph.EP},
       adsurl = {https://ui.adsabs.harvard.edu/abs/2023AJ....166..182L}
}

@ARTICLE{2026ApJ...999..210G,
       author = {{Gajjar}, Vishal and {Brown}, Grayce C.},
        title = "{Exo-IPM Scattering as a Hidden Gatekeeper of Narrowband Technosignatures}",
      journal = {\apj},
         year = 2026,
        month = mar,
       volume = {999},
       number = {2},
          eid = {210},
        pages = {210},
          doi = {10.3847/1538-4357/ae3d33},
       adsurl = {https://ui.adsabs.harvard.edu/abs/2026ApJ...999..210G}
}


\end{CJK*}
\end{document}